\documentclass[12pt]{article}                   
\usepackage{amssymb}
\usepackage{amsmath}
\usepackage{verbatim}
\usepackage{graphicx} 
\usepackage{cite}
\usepackage{geometry}             
\usepackage[hyperindex=true,
          pdfstartview=FitH,
          bookmarksnumbered=true,
          bookmarksopen=true,
          citecolor=blue,
          linkcolor=blue,
          colorlinks=true,
          pdfborder=001,
          unicode]{hyperref}

\newcommand{\rd}{\mathrm{d}}
\newcommand{\pfrac}[2]{\left(\frac{\partial #1}{\partial #2}\right)}

\allowdisplaybreaks

\newcommand{\blue}[1]{\textcolor{blue}{#1}}
\renewcommand{\blue}[1]{\textcolor{black}{#1}}  

\begin{document}

\title{Black hole thermodynamics is extensive \\
with variable Newton constant}
\author{Tao Wang and Liu Zhao\thanks{Correspondence author.}\\
School of Physics, Nankai University, Tianjin 300071, China
\\
{\em email}: \href{mailto:taowang@mail.nankai.edu.cn}
{taowang@mail.nankai.edu.cn}
and
\href{mailto:lzhao@nankai.edu.cn}{lzhao@nankai.edu.cn}
}

\date{}

\maketitle

\begin{abstract}
Inspired by the recent studies on the thermodynamics of AdS black holes 
in the restricted phase space formalism, we propose a similar formalism for 
the thermodynamics of non-AdS black holes with variable Newton constant.
It is shown that, by introducing the new variables $N,\mu$, where $N$ is proportional
to the inverse Newton constant and $\mu$ its conjugate variable, referred to as the 
chemical potential, the  black hole thermodynamics can be formulated in a form which 
is consistent with the standard extensive thermodynamics for
open macroscopic systems, with the first law and 
the Euler relation hold simultaneously. This formalism has profound implications,
in particular, the mass is a homogeneous function of the first order in the extensive
variables and the intensive variables are zeroth order homogeneous functions. The 
chemical potential is shown to be closely related to the Euclidean action 
evaluated at the black hole configuration.

\end{abstract}

\section{Introduction\label{Intro}}

In connection with the recent studies \cite{Zhao1,Zhao2}
on the thermodynamics of AdS black holes in 
the reduced phase space (RPS) formalism, which indicated that the thermodynamics
for AdS black holes is extensive in the standard sense, e.g. with the correct Euler relation 
and homogeneity behaviors, it is natural to ask what about the cases for 
non-AdS black holes. Since the newly introduced 
variables $C,\mu$ known as the central charge and the chemical potential 
respectively in the RPS formalism rely on the holographic interpretation 
\cite{Visser} (see also \cite{Kastor2,Zhang,Karch,Maity,Wei}), the first 
impression may be that this 
formalism should work only for AdS black holes. The purpose of this work is to 
acknowledge that a similar formalism also works for asymptotically
flat and de Sitter (dS) black holes, albeit the lack of holographic interpretation.
Since there is no extended phase space formalism in these cases, it may not be 
appropriate to refer to the corresponding formalism as RPS thermodynamics.
Therefore we simply call this new formalism as thermodynamics with variable 
Newton constant. 

Extensivity is a much needed and long awaited feature for black hole 
thermodynamics which has not been realized before the works \cite{Zhao1,Zhao2}.
In the absence of extensivity, the previous formalisms for black hole 
thermodynamics such as the traditional \cite{Bekenstein,Bekenstein2,Bardeen,Hawking}
and the extended phase space \cite{Kastor,Dolan,Dolan2,Dolan3,Kubiznak,Cai,Kubiznak2} 
formalisms suffer from a number of issues, e.g. \blue{the black hole mass is not a
homogeneous function in the extensive variables of the first order \cite{G,York,Gibbons2} 
as required in standard textbooks in thermodynamics, see, e.g. \cite{Callen}; 
the interpretation for the total mass as enthalpy 
in the extended phase space formalism \cite{Kastor}
seems to be in contradiction with the standard understanding of AD(M) mass as the 
conserved charge associated with the timelike Killing symmetry, {\em etc.} Moreover,}
some of the exotic thermodynamic behaviors such as intersecting isotherms and 
\blue{discontinuous Gibbs free energies} observed in the extended phase space formalism
\blue{do not appear in extensive description of thermodynamic systems.}
It is important to stress that, pointing out the issues 
in existing formalisms does not disvalue the previous works. On the contrary, 
there is nothing wrong in the previous works, the issues just indicate that there are some 
missing dimensions in the space of macro states for the black holes, which
are identified to be the central charge and the chemical potential in the AdS cases.
The clue for identifying the missing dimensions for the asymptotically flat and dS 
cases has already appeared in previous works, it only needs some time and insight 
to spell it out correctly, that is when and for what we come in.


\section{RPS formalism for AdS and the logic behind}

The RPS formalism for the AdS cases is derived from Visser's 
holographic thermodynamics involving an AdS black hole spacetime and a dual CFT.
The basic structure of Visser's framework is captured by the following two equations,
\begin{align}
\rd E &= T\rd S -\mathcal{P}\rd\mathcal{V}+\tilde \Phi\rd\tilde Q+\Omega\rd J
+\mu\rd C,\label{1st}\\
E&=TS +\tilde \Phi\tilde Q+\Omega J+\mu C,
\label{Elike}
\end{align}
where $E,T,S,\Omega,J$ bear the usual sense as in literally any related references,
$\tilde \Phi,\tilde Q$ are respectively the electric potential and charge with some proper
rescaling, and $C,\mu$ are respectively the central charge and the chemical potential.
The variables $\mathcal{P},\mathcal{V}$ are defined such that $\mathcal{V}\sim L^{d-2}$ 
represents the spatial volume of the CFT and $\mathcal{P}$ is determined by the 
caloric equation of states $E=(d-2)\mathcal{PV}$ of the CFT, 
where $L$ is the AdS radius and $d$ the 
dimension of the bulk spacetime. The RPS formalism arises by fixing the AdS radius $L$,
which effectively eliminates the variables $\mathcal{P},\mathcal{V}$,
yielding
\begin{align}
\rd E &= T\rd S +\tilde \Phi\rd\tilde Q+\Omega\rd J
+\mu\rd C. \label{1st2}
\end{align}
Eq.\eqref{1st2} represents the first law and \eqref{Elike} gives the corresponding 
Euler relation in the RPS formalism. Notice that eq.\eqref{Elike} is {\em not} the 
correct Euler relation associated with eq.\eqref{1st}, because there is a missing 
term $- \mathcal{P}\mathcal{V}$.

The logic behind the construction is as follows. The central charge $C$ encodes the 
level of conformal anomaly in the dual CFT, which takes the value
\begin{align}
C=\frac{L^{d-2}}{G},	\label{central}
\end{align}
and is related to the number of colors $\mathcal{N}$ via $C \propto \mathcal{N}^2$ 
or $C\propto \mathcal{N}^{3/2}$ 
depending on the underlying gravity model \cite{Rafiee:2021hyj}. \blue{While both $L$ 
and $G$ are considered to be variable in \cite{Visser}, the RPS formalism keeps $L$ fixed
and only $G$ is allowed to vary. Physically, $G$ is proportional to $(\ell_{\rm P})^{d-2}$,
where $\ell_{\rm P}$ represents the Planck length. Therefore, $C$ has an 
intuitive interpretation as the number of pieces of the size of Planck ``area'' that a 
hypersurface of radius $L$ can be divided into, with each such piece representing a single
microscopic degree of freedom. Varying $G$ implies changing the 
Planck length, which in turn changes the number of pieces mentioned above.}
The CFT resides on the boundary of the black hole spacetime, of which the metric 
can be obtained using a limiting procedure. For instance, assuming that the bulk spacetime 
has the line element
\[
\rd s^2_{\rm bulk}=-A(r)\rd t^2+\frac{1}{A(r)}\rd r^2+\cdots+r^2\rd\Omega^2_{d-2}
\]
where $A(r)$ has the asymptotic behavior
\[
A(r)\sim \frac{r^2}{L^2}, \quad \mbox{ as } \quad{r\to\infty},
\]
the boundary metric is obtained via
\[
\rd s^2_{\rm bdry} = \lim\limits_{r \to \infty}\frac{R^2}{r^2}\,\rd s^2_{\rm bulk}
=-\rd\tau^2+{R^2}\rd\Omega^2_{d-2},
\]
where $R$ is some large constant with the dimension of $L$. The crucial point lies in that,
setting $R=L$, the thermodynamic quantities for the bulk black hole are precisely identical
to those of the dual CFT, e.g. 
\[
T_{\rm bulk}= T_{\rm CFT},\quad S_{\rm bulk}= S_{\rm CFT},\quad etc.
\]
One can establish further relationships between the two theories using the AdS/CFT 
dictionaries, of which the most basic one is the identification of the 
partition functions on both sides,
\begin{align}
Z_{\rm CFT}=Z_{\rm gravity} =\exp(-\mathcal{A}_{E}),\label{ZZ}
\end{align}
where $\mathcal{A}_{E}$ is the Euclidean action in the bulk taken at the black hole 
configuration. 

An important observation is that the Gibbs free energy $W$ of the CFT 
can be defined via 
\[
W \equiv -T\log Z_{\rm CFT}=\mu C,
\]
which, by use of the AdS/CFT dictionary \eqref{ZZ}, leads to
\begin{align}
W =\mu C = T \mathcal{A}_{E}, \qquad 
\mu =\frac{T\mathcal{A}_{E}}{C}= \frac{G T \mathcal{A}_{E}}{L^{d-2}} .
\label{mu}
\end{align}
Since the thermodynamic quantities for the bulk and the boundary theories are identical, 
$\mu$ and $C$ are also understood to be thermodynamic quantities on the black hole side.
We thus see that the chemical potential for the black hole can be defined independent 
of the Euler relation \eqref{Elike}, and the Euclidean action plays an 
indispensable role. With $W=\mu C$ in mind, the work presented in \cite{Gibbons2}
implies the correctness of the Euler relation for AdS black holes.

\section{Thermodynamics with variable Newton constant}

The discussion made in the last section has borrowed two concepts from the AdS/CFT
correspondence, i.e. the central charge and the chemical potential. For non-AdS
black holes one encounters an immediate difficulty due to the lack of the CFT 
description. In particular, the definition of the central charge relies on 
the AdS radius $L$, a characteristic length scale which is absent in asymptotically 
flat spacetimes.

However, as we will show below, the parallel line of thinking still works 
for non-AdS black holes. The only change is to replace $L$ by 
some constant length scale, also denoted as $L$, with completely different meaning. 
For asymptotically flat black holes, $L$ may be regarded as
the radius of a containing box\footnote{\blue{By a containing box, we refer to 
a hypothetical spatial hypersurface which, besides marking the largest possible scale of the 
event horizon, has no physical consequences and represents no physical reality. 
This concept should be distinguished from {\em confining box} or {\em cavity}, 
which represents a physical boundary which has extra consequences, e.g. in
modifying the total stress energy tensor \cite{Lemos1}.}}, 
which corresponds to the largest possible radius 
which the event horizon can reach during the thermodynamic process of
interests.

To check this idea, let us consider the simplest example of four dimensional 
asymptotically flat Schwarzschild black hole with metric
\begin{align}
\rd s^2&=-f(r)\rd t^2+ f(r)^{-1}\rd r^2
+r^2\left(\rd\theta^2+\sin^2\theta\rd\phi^2\right),\label{metric}\\
f(r)&=1-\frac{2GM}{r}.
\end{align}
The event horizon is located at $r=r_g=2GM$, and the temperature
observed by static observers located at spatial infinity and the entropy are 
respectively given by 
\begin{align}
T=\frac{f'(r_g)}{4\pi}=\frac{1}{8\pi GM},\quad 
S=\frac{A}{4G}=\frac{\pi r_g^2}{G}={4\pi GM^2}	.
\end{align}
The traditional formalism treats $G$ as constant and hence 
\begin{align}
\rd M = T\rd S,\quad M=2TS.	 \label{MSmar}
\end{align}
This is the first law and the famous Smarr relation \cite{Smarr}.

Now let us proceed by taking $G$ as a variable. Assuming that the black hole is 
contained in a spherical box of radius $L$ centered at $r=0$.
Due to the fact that the event horizon
is a causal boundary, all meaningful degrees of freedom (whatever they are) must
be distributed on the horizon surface, rather than in the volume surrounded by 
the horizon. Actually, there is no such notion as the volume surrounded by 
the horizon. Therefore, it is reasonable to assume that the number $N$ of microscopic 
degrees of freedom for the black hole is proportional to $L^2$, a measure of the largest 
possible area which the event horizon can reach during the thermodynamic process of 
interests. In particular, we assume that 
\begin{align}
N=\frac{L^2}{G}.  \label{N}
\end{align}
\blue{Although $N$ does not have the holographic interpretation as the central charge, 
it still represents the number of pieces of the size of Planck area 
that a spherical box of radius $L$ can be divided into.}
We also introduce the chemical potential $\mu$ {\em by use of the Euler relation}
\begin{align}
M=TS+\mu N, \label{Euler2}
\end{align}
which ensures that $\mu, N$ form a conjugate pair. Explicitly, we have, for Schwarzschild
black hole,
\begin{align}
\mu 
=\frac{G M}{2L^2}.	
\end{align}
Considering $S$ and $N$ as functions of $M$ and $G$, we can easily check that 
\begin{align}
\rd M=T\rd S+\mu \rd N.
\label{fst}
\end{align}
This gives a simple example for black holes in asymptotically flat spacetime 
in which the first law and the Euler relation 
hold simultaneously with variable Newton constant. It is interesting to notice that,
in the above construction, $L$ played no essential role and hence it can have any 
constant value. However, by dimensional consideration it is necessary to introduce 
a length scale $L$ in eq.\eqref{N}, rather than simply defining $N$ by the 
inverse of $G$. \blue{Please also note that even when $G$ is kept fixed and the 
first law \eqref{fst} reduces into the traditional form presented in eq.\eqref{MSmar},
the variables $N$ and $\mu$ are still meaningful, the Euler relation \eqref{Euler2} 
still holds and is distinct from the Smarr relation. One can think of our formalism
with fixed $G$ as a thermodynamic description for the black hole as a closed system,
while the case with variable $G$ as a description of the black hole as an open system.}

We can go a little further
by writing $M$ as a first order homogeneous function in $S$ and $N$. This is achieved by
solving $M$ from the expression for $S$ and inserting the definition of $N$ in the result.
The final expression is given as
\begin{align}
M(S,N)= \sqrt{\frac{S}{4\pi G}}=\frac{1}{2L}\sqrt{\frac{NS}{\pi}},
\label{MSN}
\end{align}
which is clearly a first order homogeneous function in $S$ and $N$. For comparison,
in the traditional formalism with fixed Newton constant, the mass is 
\blue{a homogeneous function 
in $S$ of order one half}, thanks to the Smarr relation \eqref{MSmar}, which
looks strange if $M$ is to be taken as a thermodynamic potential as it should be.

Eq.\eqref{MSN} allows us to get the equations of states
\begin{align}
T&=\pfrac{M}{S}_N= \frac{1}{4L}\sqrt{\frac{N}{\pi S}},\\
\mu&=\pfrac{M}{N}_S = \frac{1}{4L}\sqrt{\frac{S}{\pi N}}. 
\label{MUSN}
\end{align}
Using these equations we can explicitly verify the extensivity of $S$ by rewriting
\[
S=\frac{N}{16\pi L^2 T^2} = 16\pi L^2\mu^2 N,
\]
i.e. $S$ is proportional to $N$ with the coefficient of proportionality depending 
only on the intensive variables.

We can also introduce the heat capacity for Schwarzschild black hole unambiguously,
\[
C_N=T\pfrac{S}{T}_N = -\frac{N}{8\pi L^2 T^2}=-2S.
\]
As expected, $C_N$ is always negative, which signifies the thermodynamic instability of
the Schwarzschild black hole. 

The surprising validity for the first law and Euler relation in the Schwarzschild 
case with variable $G$ forces us to consider also the validity of the 
same formalism for de Sitter black holes. To this end, let us again consider 
a simple example, i.e. Schwarzschild-de Sitter case. The metric is still written in 
the form \eqref{metric}, but with $f(r)$ replaced by
\begin{align}
f(r)=1-\frac{2GM}{r}-\frac{\Lambda r^2}{3},	
\end{align}
where the cosmological constant $\Lambda>0$ corresponds to de Sitter case and 
$\Lambda<0$ corresponds to AdS case. \blue{We assume that $9\Lambda G^2M^2<1$, so that
$f(r)$ has two distinct real zeros $r_h$ and $r_c$ which correspond respectively to the 
black hole event horizon and the cosmological horizon.}

\blue{Comparing to the cases of asymptotically flat and AdS black holes, the 
thermodynamics for de Sitter black holes is more subtle due to the presence of two horizons. 
The existence of the cosmological horizon makes it impossible to normalize the 
timelike Killing vector at spatial infinity, which brings in some ambiguity in 
determining the values of surface gravity and hence the temperatures. 
A reasonable way out is to consider thermodynamic relations on only one of the horizons, 
fixing the other as a boundary \cite{Gomberoff}. Then the temperature can be determined 
unambiguously using the Euclidean period approach. 
In the following, we shall follow this line of thinking and consider only the 
thermodynamics associated with the event horizon. 
Similar constructions associated with the cosmological horizon should also work, 
but we simply omit it for simplicity and brevity.}

Now $M$ can be represented in 
terms of the radius of the event horizon $r_h$ by use of the equation
$f(r_h)=0$ which determines the horizon radius, i.e.
\begin{align}
M= \frac{r_h(3-\Lambda r_h^2)}{6G}.	 \label{MdS}
\end{align}
The associated temperature and entropy are given by 
\begin{align}
T&=\frac{f'(r_h)}{4\pi}= \frac{2 G M}{r_h^2}-\frac{2 \Lambda r_h}{3}
=\frac{1-\Lambda r_h^2}{4\pi r_h},\\
S&= \frac{A}{4G}=\frac{\pi r_h^2}{G}.
\end{align}
Notice that both $M$ and $S$ are inversely proportional to $G$ if they are considered 
as functions of $(r_h,G)$. 
This feature will be useful in later analysis.

The number $N$ of microscopic 
degrees of freedom for the black hole is still assumed to take the form \eqref{N}, 
however please be aware that $L$ is not related in any way to the cosmological constant.
Using the Euler relation $M=TS+\mu N$, we can calculate the value of $\mu$,
yielding
\begin{align}
\mu= \frac{r_h(3+\Lambda r_h^2)}{12L^2}. \label{mudS}
\end{align}
Finally, it is straightforward to check that the first law \eqref{fst} indeed holds 
in this case, assuming that $G$ is an independent variable.

From the values for $N$ and $S$ we can get
\[
G=\frac{L^2}{N},\quad r_h=L\sqrt{\frac{S}{\pi N}}.
\]
Inserting these relations into eqs.\eqref{MdS}-\eqref{mudS} yields
\begin{align}
M&=	\frac{3 \pi N-\Lambda  L^2 S }{6 \pi ^{3/2} L} \sqrt{\frac{S}{N}}, 
\label{msn}\\
T&= \frac{\pi N-\Lambda  L^2 S}{4 \pi ^{3/2} L S} \sqrt{\frac{S}{N}},\\
\mu&= \frac{3 \pi N+\Lambda  L^2 S}{12 \pi ^{3/2} L N} \sqrt{\frac{S}{N}}.
\label{musn}
\end{align}
The first order homogeneity of $M$ and zeroth order homogeneity of $T,\mu$ in 
$S,N$ are crystal clear from the above equations. 

\blue{Although eqs.\eqref{msn}-\eqref{musn} are derived specifically for the de Sitter case,
they actually hold also for asymptotically flat and AdS cases. In particular, 
for $\Lambda=0$, eqs.\eqref{msn}-\eqref{musn} fall back to eqs. \eqref{MSN}-\eqref{MUSN}; 
for $\Lambda<0$, $L$ can be chosen to be independent of the AdS radius, which is a feature 
not mentioned in our previous work \cite{Zhao1}.}

\blue{Let us now proceed with some discussions for generic choice of $\Lambda$.}
In order to ensure $T\geq 0$, we need
\begin{align}
\pi N-\Lambda L^2 S\geq 0.
\label{bd}
\end{align}
The heat capacity evaluated using the above equations of states reads
\[
C_N= - 2S\left(\frac{\pi N-\Lambda  L^2 S}{\pi N+\Lambda  L^2 S}\right).
\]
For $\Lambda\to 0$, we have $C_N\to -2S$, which recovers the Schwarzschild result. For 
$\Lambda>0$, $C_N$ is always non-positive, thanks to the bound
\eqref{bd}, which indicates that the Schwarzschild-de Sitter black hole is 
also thermodynamically unstable. For $\Lambda <0$, the bound \eqref{bd} 
imposes no restriction on the $N$ and $S$ values, thus for sufficiently large $S$ 
which makes $\pi N+\Lambda  L^2 S<0$ (with $\Lambda<0$), 
$C_N$ can become positive. Therefore, the Schwarzschild-AdS black hole can 
be thermodynamically stable. 

Another point to be noticed is that, according to eq.\eqref{musn}, the chemical potential
is strictly positive if $\Lambda\geq 0$, which implies that the microscopic degrees of 
freedom are repulsive, which might help to understand the thermodynamic instability
in these cases. On the contrary, if $\Lambda<0$, $\mu$ becomes zero at 
$N=-\frac{\Lambda L^2 S}{3\pi}$ and negative for $N<-\frac{\Lambda L^2 S}{3\pi}$. 
The former condition reconfirms that the Hawking-Page 
transition \cite{Hawking2} appears only for AdS but not for asymptotically 
flat and de Sitter black holes, while the latter signifies that the microscopic 
degrees of freedom are attractive for sufficiently large $S$, 
and hence the large AdS black hole becomes thermodynamically stable.

The proceeding two simple examples make us confident that the same formalism 
involving variable $G$ should work for generic black holes, be it AdS or not. 
The correct Euler relation is a key ingredient in this formalism, which guarantees 
the extensivity. Indeed, we have verified the validity of eq.\eqref{1st2} for Kerr-Newman 
black holes in both asymptotically flat and de Sitter spacetimes, 
where $C$ is given by eq.\eqref{central} and $\mu$ is defined via the Euler 
relation \eqref{Elike}\footnote{To keep the presentation of the present manuscript 
as clean as possible, we omit the details for the Kerr-Newman and Kerr-Newman-de Sitter
cases. The readers are encouraged to reproduce the explicit procedures by themselves,
which is a little involved but can still be carried out using simple algebra and calculus.}. 
There is no necessary relationship between $L$ and $\Lambda$, 
the only requirement is that both of them are constant, and that $G$ is treated as 
a variable which defines $C$. That the Euler relation and the first law together guarantees 
extensivity of the thermodynamics can be explained very easily. We can simply
rewrite eq.\eqref{Elike} in the form (we use $M$ in place of $E$ and $N$ in place of $C$)
\begin{align}
M=\pfrac{M}{S}_{\tilde Q,J,N} S+\pfrac{M}{\tilde Q}_{S,J,N}\tilde Q
+\pfrac{M}{J}_{S,\tilde Q,N}J+\pfrac{M}{N}_{S,\tilde Q,J}N.
\end{align}
Then it is clear that $M$ scales as $M\to \lambda M$ if $S,\tilde Q,J,N$ all scales 
in the same way, i.e. $S\to \lambda S$, {\em etc}. Then it is evident that the partial 
derivatives, which are actually the intensive variables $T,\hat \Phi, \Omega,\mu$,
will not get rescaled. What remains is a proper understanding about $\mu$, 
because now it is defined using the Euler relation, rather than defined 
independently as in the AdS cases.

\section{Understanding the chemical potential}

In the variable $G$ formalism discussed in the last section,
the chemical potential $\mu$ is defined via the Euler relation. This is somewhat puzzling
and needs further understanding, because the first law and the Euler relation are
commonly regarded as independent relations. 

To solve the puzzle, let us recall that, for Kerr-Newman black hole, the Euclidean action 
$\mathcal{A}_E$ is verified to satisfy the relation \cite{G}
\[
W=T \mathcal{A}_E = M- TS- \Phi Q-\Omega J.
\]
This is a purely algebraic relation which does not rely on whether $G$ 
is variable or not. The above relation reduces to 
\begin{align}
T \mathcal{A}_E = M- TS 
\label{euca}
\end{align}
in the case of static neutral black holes.
Meanwhile, for fixed $G$, the traditional formalism of black hole thermodynamics 
admits the first law
\[
\tilde{\rd} M =T\tilde{\rd} S +\Phi\tilde{\rd} Q+\Omega\tilde{\rd}J,
\]
which reduces to 
\[
\tilde{\rd} M =T\tilde{\rd} S
\]
for static neutral black holes, 
where $\tilde{\rd}$ represents the total differential taken when $G$ is 
considered to be a constant.

Now if $G$ allowed to vary, then
\begin{align}
\rd M =\tilde\rd M+ \pfrac{M}{G} \rd G
=\tilde\rd M-M\frac{\rd G}{G},  \label{ddtM}
\end{align}
where we have used the fact that $M$ is inversely proportional to $G$. 
Similarly we have
\[
\rd S =\tilde\rd S - S \frac{\rd G}{G}.
\]
Using the above relations we have
\begin{align}
\rd M &= T\left[\rd S+ S\frac{\rd G}{G}\right]-M\frac{\rd G}{G}
=T\rd S - (M-TS)\frac{\rd G}{G}\nonumber\\
&= T\rd S + \frac{M-TS}{N}\rd N = T\rd S + \mu\rd N.
\end{align}
We recognize that this is precisely the first law in the formalism with variable $G$, 
in which $\mu$ is given by use of the Euler relation \eqref{Euler2}.
The same procedure also works for more general black hole solutions in generic 
dimension $d$. However, please be reminded that, for charged black holes, $M$ is
not inversely proportional to $G$, because there is a $G$ factor in front of $Q^2$
which appears in the expression for $M$. In order to make sure that the first law 
still holds with variable $G$, a $G$-dependent rescaling of $Q$ and $\Phi$ 
maintaining the product $\Phi Q$ is needed,
as did in \cite{Zhao1}. 

Comparing the Euler relation \eqref{Euler2} with 
the Euclidean action formula \eqref{euca}, we get
\[
\mu= \frac{T\mathcal{A}_E}{N}=\frac{G T\mathcal{A}_E}{L^{2}}.
\]
This reproduces the result of eq.\eqref{mu} for $d=4$, but now for non-AdS 
black holes without holographic interpretation. \blue{The fact that the chemical potential 
is connected with the Euclidean action is natural. Let us remark that, since the 
Newton constant is an overall factor in the Einstein-Hilbert action (denoted 
as $\mathcal{A}$), 
the expression $G \mathcal{A}/L^2$ is the {\em mechanical} conjugate to $N=L^2/G$. 
The thermodynamic conjugate differs from the mechanical conjugate in that the 
temperature $T$ enters in the expression for $\mu$, and that the action 
$\mathcal{A}$ is replaced by its Euclidean counterpart $\mathcal{A}_E$.}

Before closing this section, let us mention that the Euclidean action for RN-AdS
black hole together with the $\Lambda\to 0$ limit was considered in \cite{Lemos1}. 
The connection between Euclidean action approach and the extended phase space formalism was 
discussed in \cite{Lemos2}. In either cases the Euclidean action is not associated with 
the chemical potential and the formalisms are thus not extensive.

\section{Concluding remarks}

The work presented in this paper indicates that the thermodynamics for black holes 
fall in the standard framework of extensive thermodynamics, provided the Newton 
constant is allowed to vary. This is true for all kinds of asymptotics, 
including asymptotically flat and de Sitter cases. Although the latter cases 
do not admit holographic interpretations, the relation
$\mu= \frac{T\mathcal{A}_E}{N}$ between the chemical potential $\mu$ 
and the Euclidean action $\mathcal{A}_{E}$ 
is universal. We hope the new formalism will be useful in further exploring and 
understanding the physics of black holes from the macroscopic perspective. 

While finishing this manuscript, we noticed that the idea of including variable 
Newton constant in the formulation of black hole thermodynamics has already 
been used in \cite{Volovik}. However, the motivation and the major points 
of interests are different from ours. In particular, extensivity of the formalism,
including the Euler relation and the related homogeneity behaviors
are not considered in \cite{Volovik}.

\section*{Acknowledgement}
This work is supported by the National Natural Science Foundation of 
China under the grant No. 11575088.

\providecommand{\href}[2]{#2}\begingroup
\footnotesize\itemsep=0pt
\providecommand{\eprint}[2][]{\href{http://arxiv.org/abs/#2}{arXiv:#2}}

\end{document}